\documentclass[showpacs, preprintnumbers, amsmath, amssymb, floatfix]{revtex4-2}
\usepackage{graphicx}
\usepackage{dcolumn}
\usepackage{bm}
\usepackage{amsmath,amsthm,amssymb,ascmac}
\usepackage{color}

\newcommand{\g}{{g \hspace{-1.6mm}^{^\circ}}}
\newcommand{\gu}{{{g^{ab}} \hspace{-4.5mm}^{^\circ} \ \ }}
\newcommand{\be}{\beta}
\newcommand{\ka}{\kappa}
\newcommand{\aeq}{&=}

\newcommand{\aeqd}{&:=}
\newcommand{\mL}{\mathcal{L}}

\newcommand{\mG}{\mathcal{G}}
\newcommand{\mF}{\mathcal{F}}
\newcommand{\mK}{\mathcal{K}}
\newcommand{\G}{{\rm{G}}}
\newcommand{\e}{\equiv}
\newcommand{\ep}{\varepsilon}

\newcommand{\al}{\alpha}

\newcommand{\f}{\frac}
\newcommand{\half}{\frac{1}{2}}
\newcommand{\pr}{\prime}
\newcommand{\dl}{\delta}
\newcommand{\Dl}{\Delta}

\newcommand{\om}{\omega}
\newcommand{\Om}{\Omega}
\newcommand{\com}{\hspace{0.5mm}, \quad}
\newcommand{\la}{\label}
\newcommand{\no}{\nonumber}
\newcommand{\re}[1]{(\ref{#1})}
\newcommand{\res}[1]{\S \ref{#1}}

\newcommand{\RM}[1]{{\rm{#1}}}
\newcommand{\p}{\partial}
\newcommand{\w}{\wedge}
\newcommand{\co}[1]{``{#1}''}
\newcommand{\bea}[1]{\begin{align} 
#1
\end{align}}
\newcommand{\rd}[1]{\textcolor{black}{#1}}

\begin{document}
\title{Noether currents and generators of local gauge transformations in the covariant canonical formalism}

\author{Satoshi Nakajima}
\email{subarusatosi@gmail.com}

\date{\today}

\begin{abstract}
We investigate generators of local transformations in the covariant canonical formalism (CCF). 
The CCF treats space and time on an equal footing regarding the differential forms as the basic variables.
The conjugate forms $\pi_A$ are defined as derivatives of the Lagrangian $d$-form $L(\psi^A, d\psi^A)$ with respect to $d\psi^A$, namely $\pi_A := \partial L/\partial d\psi^A$, 
where $\psi^A $ are $p$-form dynamical fields.
The form-canonical equations are derived from the form-Legendre transformation of the Lagrangian form $H:=d\psi^A \wedge \pi_A - L$.
We show that the Noether current form is the generator of an infinitesimal transformation $\psi^A \to \psi^A + \delta  \psi^A$
if the transformation of the Lagrangian form is given by $\delta L=dl$ and $\delta  \psi^A$ and $l$ depend on only $\psi^A$ and the parameters. 
As an instance, we study the local gauge transformation for the gauge field and the local Lorentz transformation for the second order formalism of gravity. 

\end{abstract}

\maketitle

\section{Introduction}
In the classical field theory, the traditional canonical formalism gives special weight to time.
The covariant canonical formalism (CCF) \cite{10, 11, 12, 13, N04, K, 2015, 2016, 2017, 2019, 2020} is a covariant extension of the traditional canonical formalism.
The form-Legendre transformation and the form-canonical equations are derived from a Lagrangian $d$-form with $p$-form dynamical fields $\psi^A$. 
The conjugate forms are defined as derivatives of the Lagrangian form with respect to $d\psi^A$. 
One can obtain the form-canonical equations of gauge theories or those of the second order formalism of gravity without fixing a gauge 
nor introducing the Dirac bracket or  any other artificial tricks.
Although the second order formalism of gravity (of which the dynamical variable is only the frame form (vielbein)) 
is a non-constrained system in the CCF, the first order formalism (of which the dynamical variables are both the frame form and the connection form) 
is a constrained system even in the CCF. 
In Refs.\cite{10,11,12,13}, the CCFs of the first order formalism of gravity and supergravity have been studied. 
The CCF of the second order formalism of gravity without Dirac field \cite{K} and with Dirac field \cite{2015} have been studied. 

Poisson brackets of the CCF are proposed in Refs.\cite{10, 2019} and in Ref.\cite{2017} independently.
These are equivalent. 
Although the form-canonical equations of the CCF are equivalent to modified De Donder-Weyl equations \cite{2016}, 
the Poisson bracket of the CCF is not equivalent to it of the De Donder-Weyl theory \cite{IV, DW2021}. 
Reference \cite{2019} introduced the generator of the CCF and studied it of the local Lorentz transformation of gravity in the first order formalism. 
The generators of the local Lorentz transformation and the supersymmetry for supergravity have been studied \cite{2020} in the first order formalism. 
However, relations between the Noether current and the generator of the CCF were not clear.

The structure of the paper is as follows.
First, we review the covariant canonical formalism (\res{s_rev_CCF}).
Next, in \res{s_Noether}, we consider the Noether current form $N$ for an infinitesimal local transformation $\psi^A \to \psi^A + \dl \psi^A$, $\rd{\dl L}=dl$ 
where $\psi^A$ are differential forms of the dynamical fields and $L$ is the Lagrangian form. 
If $\dl \psi^A$ and $l$ depend on only $\psi^A$ and the parameters, the Noether current $N$ is the generator of the transformation of the CCF (\res{s_main}). 
As an instance, we study the local gauge transformation for the gauge field coupled with a matter field (\res{s_gauge}) 
and the local Lorentz transformation for the second order formalism of gravity coupled with Dirac fields (\res{s_Gravity_D}). 
Our demonstrations are for non-constrained systems, in contrast to previous studies for constraint systems.
\rd{In Appendix \ref{A_PB}, we derive \re{pb}.}
\rd{In Appendix \ref{YMU}, we review the CCF for the gauge field.}
In Appendix \ref{formula}, several formulas are listed.
In Appendix \ref{A_derivation}, we derive \re{F_ab}.
\rd{In Appendix \ref{CCF_gravity}, we study the canonical equations of the second order formalism of gravity in the arbitrary dimension. }

\section{Covariant canonical formalism} \la{s_rev_CCF}

In this section, we review the covariant canonical formalism in $d$ dimension space-time.
 
Suppose a $p$-form $\be$ depends on forms $\{\al^i \}_{i=1}^k$. 
If there exists the form $\om_i$ such that $\be$ behaves under variations $\dl \al^i$ as 
$\dl \be = \dl \al^i \w \om_i $, we call $\om_i$ the {\it derivative} of $\be$ with respect to $\al^i$ 
and denote 
\bea{
\f{\p \be}{\p \al^i} := \om_i.
}
In this case, $\be$ is {\it differentiable} with respect to $\al^i$.

The Lagrangian $d$-form $L$ is given by $L=\mL \eta$ where $\mL$ is the Lagrangian density and $\eta= \ast 1$ is the volume form 
($\ast$ is the Hodge operator) and depends on $\psi$ and $d\psi$, $L=L(\psi,d\psi)$, where $\psi$ is a set the forms of the dynamical fields.
For simplicity, we treat $\psi$ as single $p$-form in this section. 
The Euler-Lagrange equation is given by
\bea{
\f{\p L}{\p \psi}-(-1)^p d\f{\p L}{\p d\psi} = 0 .\la{EL}
}
The above Euler-Lagrange equation has been used since the 1970's \cite{72, 78, 95}.

The {\it conjugate form} $\pi$ is defined by
\bea{
\pi := \f{\p L}{\p d\psi}.
}
This is a $q$-form where $q:= d-p-1$. 
The {\it Hamilton $d$-form} is defined by
\bea{
H(\psi,\pi):= d\psi \w \pi-L 
}
and depends on $\psi$ and $\pi$. 
The variation of $H$ is given by
\bea{
\dl H = (-1)^{(p+1)q}\dl \pi \w d\psi -\dl \psi \w\f{\p L}{\p \psi}.
}
Thus, we obtain
\bea{
\f{\p H}{\p \psi}=  -\f{\p L}{\p \psi}\com
\f{\p H}{\p \pi}= (-1)^{(p+1)q}d\psi .
}
By substituting the Euler-Lagrange equation \re{EL}, we obtain the {\it canonical equations}
\bea{
d\psi  = (-1)^{(p+1)q}\f{\p H}{\p \pi} \com d\pi = -(-1)^p\f{\p H}{\p \psi}.
}

The {\it Poisson bracket} proposed in Ref.\cite{2017} is given by
\bea{
\{A,B \} = (-1)^{p(a+d+1)} \f{\p A}{\p \psi}\w \f{\p B}{\p \pi}-(-1)^{(d+p-1)(a+1)}\f{\p A}{\p \pi}\w \f{\p B}{\p \psi}. \la{PB}
}
Here, $A$ and $B$ are differentiable with respect to $\psi$ and $\pi$, and $A$ is an $a$-form. 
The Poisson bracket proposed in Ref.\cite{2019}, denoted by $\{A,B \}_\RM{F}$, is given by $\{A,B \}_\RM{F}=-\{B,A \}$.
If $A$, $B$ and $C$ are $a$-form, $b$-form and $c$-form respectively and differentiable with respect to $\psi$ and $\pi$, 
\bea{
\{B,A \} \aeq -(-1)^{(a+d+1)(b+d+1)}\{A,B \} ,\\
\{A,B \w C \} \aeq \{A,B \} \w C + (-1)^{(a+d+1)b}B \w \{A,C \},
}
and
\bea{
(-1)^{(a+d+1)(c+d+1)}\{A,\{B,C\}\} &+ (-1)^{(b+d+1)(a+d+1)}\{B,\{C,A\}\}\no\\
&+(-1)^{(c+d+1)(b+d+1)}\{C,\{A,B\}\}= 0
}
hold.
The  canonical equations can be written as
\bea{
d\psi  =-\{H, \psi\} \com d\pi=-\{H, \pi\} .\la{CE_PB}
}
The fundamental brackets are
\bea{
\{\psi,\pi\}=(-1)^{pd} \com \{\pi,\psi\}=-1 \com \{\psi,\psi\}=0=\{\pi,\pi\}.
}
If a form $F$ is differentiable with respect to $\psi$ and $\pi$, and $F$ does not depend on space-time points explicitly, 
\bea{
dF \aeq d\psi\w \f{\p F}{\p \psi}+d\pi\w \f{\p F}{\p \pi} \no\\
\aeq -\{H, F\} \la{dF_eom}
}
holds.

\section{Generator of local gauge transformation} \la{s_gauge_generator}

Let us consider that an infinitesimal transformation of dynamical fields $\psi^A$ and its conjugate forms $\pi_A$:
\bea{
\psi^A \to \psi^A + \dl \psi^A \com \pi_A \to \pi_A + \dl \pi_A. 
}
Here, $A$ is the label of the fields.
If there exists $(d-1)$-form $G$ such that
\bea{
\dl \psi^A = \{\psi^A, G \} \com \dl \pi_A = \{\pi_A, G \},
}
we call $G$ the {\it generator} of the transformation \cite{2019}. 
If a form $F$ is differentiable with respect to $\psi^A$ and $\pi_A$, the transformation of $F$ is given by 
$\dl F = \{F, G\} $.

In \res{s_Noether}, we review the Noether current. 
In \res{s_main}, we show that the Noether current is the generator under a few assumptions.
In \res{s_type}, we study Noether currents given by $\ep^r N_r+d\ep^r \w F_r$. 
We calculate $N_r$ and $F_r$ for a gauge field (\res{s_gauge}) and the second order formalism of gravity (\res{s_Gravity_D}).

\subsection{Noether current} \la{s_Noether}

We explain the Noether current.
For an infinitesimal transformation of $p$-form dynamical fields $\psi^A \to \psi^A + \dl \psi^A$, an identical equation
\bea{
\dl L \e \dl \psi^A \w  [L]_A +d\Big(\dl \psi^A \w \f{\p L}{\p d\psi^A} \Big) \la{delta_L_2}
}
holds. 
Here, $[L]_A := \p L/\p \psi^A-(-1)^p d(\p L/\p d\psi^A)$ and $\e$ denotes identical equation which holds without using the Euler-Lagrange equations. 
If $\dl L=dl$ holds, under the Euler-Lagrange equations $[L]_A=0$, the Noether current
\bea{
N := \dl \psi^A \w \f{\p L}{\p d\psi^A} -l \la{N_current}
}
is conserved $dN= 0$. 

\subsection{The Noether current is the generator} \la{s_main}

In the following of this paper, we suppose that $\dl \psi^A $ and $l$ depend on only $\psi^A$ and the parameters: 
These do not include $d\psi^A$ or $\pi_A$.

In this subsection, first, we calculate the transformation formula of $\pi_A$.
Next, we show that the Noether current is the generator of the transformation. 

The variation of $\psi^{\pr A}=\psi^A+\dl \psi^A$ is given by
\bea{
\Dl \psi^{\pr A} \aeq \Dl \psi^A + \Dl \psi^B \w \f{\p \dl \psi^A}{\p \psi^B} \no\\
\aeq (\dl^A_B + \xi^A_{\ B})\Dl \psi^B.\la{Dl_psi}
}
Here, we supposed that nonzero components of 
$\xi^A_{\ B} := \p \dl \psi^A/\p \psi^B$
are 0-forms.
Using \re{Dl_psi}, we have
\bea{
\Dl \psi^A \aeq (\dl^A_B-\xi^A_{\ B})\Dl \psi^{\pr B} ,\\
\Dl d\psi^A \aeq (\dl^A_B-\xi^A_{\ B})\Dl d\psi^{\pr B} -(-1)^p \Dl \psi^{\pr B} \w d\xi^A_{\ B}.
}
Here, we used $\Dl d\psi^A=d\Dl \psi^A$.
Substituting the above two equations into
\bea{
\Dl L \aeq \Dl \psi^A \w \f{\p L}{\p \psi^A}+\Dl d \psi^A \w \f{\p L}{\p d\psi^A},
}
we have
\bea{
\f{\p L}{\p d\psi^{\pr A}} \aeq (\dl^B_A-\xi^B_{\ A})\pi_B.
}
The derivative of $L^\pr=L+dl$ with respect to $d\psi^{\pr A}$ is given by
\bea{
\pi_A^\pr \aeq \f{\p L^\pr}{\p d\psi^{\pr A}} \no\\
\aeq \f{\p L}{\p d\psi^{\pr A}} +\f{\p l}{\p\psi^{\pr A}} \no\\
\aeq (\dl^B_A-\xi^B_{\ A})\pi_B+\f{\p l}{\p\psi^{ A}}.\la{pi^pr}
}

We show that the Noether current $N$ is the generator of the transformation:
\bea{
\{\psi^A,N \} \aeq (-1)^{p(p+d+1)}\f{\p N}{\p \pi_A} \no\\
\aeq \dl \psi^A ,\\
\{\pi_A,N \} \aeq -(-1)^{(d+p+1)(d-p)}\f{\p N}{\p \psi^A} \no\\
\aeq -\xi^B_{\ A}\pi_B+\f{\p l}{\p \psi^A} \no\\
\aeq \dl \pi_A.
}
Here, we used \re{pi^pr} and $\pi_A^\pr=\pi_A+ \dl \pi_A$.

\subsection{$\ep^r N_r+d\ep^r \w F_r$ type Noether current} \la{s_type}

In the following of this paper, we consider the (local) Noether currents given by
\bea{
N \aeq \ep^r N_r+d\ep^r \w F_r, \la{def_N_r_F_r}
}
where $\ep^r$ are the infinitesimal parameters (0-forms) and functions of the space-time points.
Here, $N_r$ are the (global) Noether currents for global transformation. 
Under the Euler-Lagrange equations $[L]_A=0$, $N_r$ are conserved 
$dN_r =0 $.
Using $dN=0$ and $dN_r =0 $, we have
\bea{
d\ep^r \w (N_r-dF_r) =0.
}
This leads to 
\bea{
N_r=-\{F_r, H\} .\la{G_F_relation}
}
Here, we used \re{dF_eom}.
According to Ref.\cite{2019}, \re{G_F_relation} holds without using the Euler-Lagrange equations. 
For the infinitesimal local gauge transformation for gauge field 
and the infinitesimal local Lorentz transformation for the second order formalism of gravity, 
the Noether currents are given by \re{def_N_r_F_r}. 
We calculate $N_r$ and $F_r$ for a gauge field (\res{s_gauge}) and the second order formalism of gravity (\res{s_Gravity_D}). 

\subsection{Local gauge transformation} \la{s_gauge}

We consider a gauge field coupled with $p$-form matter field $\psi^A$. 
For instance, the matter field is the scalar field. 
To simplify, we suppose that the matter field is not the Dirac field because the conjugate form of the Dirac field is not independent. 
The total Lagrangian form is given by
\bea{
L \aeq L_0(\psi^A,(D\psi)^A)+L_1, 
}
where $(D\psi)^A$ is the covariant derivative. 
Here, 
$L_0$ and $L_1$ are the Lagrangian forms for the matter field and the gauge field respectively. 
The infinitesimal gauge transformation is given by
\bea{
\dl \psi^A = \ep^r (\bm{G}_r)^A_{\ B}\psi^B  \com \dl A^r = \ep^s f^r_{\ st}A^t - d\ep^r \com \dl L_0 = 0 \com \dl L_1=0.
}
Here, $\ep^r$ are infinitesimal parameters and functions of the space-time points, $\bm{G}_r$ are representations of the generators of a linear Lie group $\mG$. 
$A^r$ is the gauge field. 
The matrices $\bm{G}_r$ satisfy 
\bea{
[\bm{G}_r, \bm{G}_s] = f^t_{\ rs}\bm{G}_t,
}
where $[A,B] := AB-BA$ and $f^t_{\ rs}$ are the structure constants of $\mG$. 
The covariant derivative is given by $(D\psi)^A := d\psi^A+A^r (\bm{G}_r)^{A}_{\ B}\w \psi^B$. 
The global Noether currents $N_r$ are given by
\bea{
N_r \aeq N_r^{(0)} +N_r^{(1)}, \\
N_r^{(0)} \aeq (\bm{G}_r)^A_{\ B}\psi^B \w \pi_A ,\\
N_r^{(1)} \aeq f^s_{\ rt}A^t \w \pi_s .
}
Here, $\pi_A$ and $\pi_r$ are conjugate forms of $\psi^A$ and \rd{$A^r$} respectively.
The global Noether currents satisfy \rd{(Appendix \ref{A_PB})}
\bea{
\{G_r , G_s \} \aeq f^t_{\ rs}G_t \ \ (G_r=N_r,N_r^{(0)}, N_r^{(1)}). \la{pb}
}
$F_r$ of \re{def_N_r_F_r} are given by
\bea{
F_r = -\pi_r.
}
We can confirm that the local Noether current $N$ is the generator of the transformation.

\rd{Note that the gauge field is a non-constrained system in the CCF (Appendix \ref{YMU}). 
 $dA^a$ can be represented by the conjugate form $\pi_a$. 
The second order formalism of gravity is also a non-constrained system in the CCF (Appendix \ref{CCF_gravity}).}

\subsection{Second order formalism of gravity} \la{s_Gravity_D}

We explain the notations.
Let $g$ be the metric of which has signature $(-+\cdots+)$, and let $\{\theta^a\}_{a=0}^{d-1}$ denote an orthonormal frame (vielbein).
We have $g=\g_{ab} \theta^a \otimes \theta^b$ with $\g_{ab}:= \RM{diag}(-1,1,\cdots,1)$.
All indices are lowered and raised with $\g_{ab}$ or its inverse $\gu$.
Let $\om^a_{\ b}$ be the connection form. 
In this paper, we suppose $\om_{ba}=-\om_{ab}$. 
The curvature 2-form $\Om^a_{\ b}$ is given by 
$\Om^a_{\ b}:= d\om^a_{\ b}+\om^a_{\ c}\w \om^c_{\ b} $.
Let $A^a_{\ b}$ be the Levi-Civita connection. 
We put $K^a_{\ b} := \om^a_{\ b} -A^a_{\ b}$.
For the infinitesimal local Lorentz transformation
\bea{
\dl \theta^a = \ep^a_{\ b}\theta^b ,\la{trans_gravity}
}
$\om^{ab}$, $A^{ab}$, and $K^{ab}$ behave as
\bea{
\dl \om^{ab} \aeq \ep^a_{\ c}\om^{cb}+\ep^{b}_{\ c}\om^{ac}-d\ep^{ab}, \\
\dl A^{ab} \aeq \ep^a_{\ c}A^{cb}+\ep^{b}_{\ c}A^{ac}-d\ep^{ab}, \\
\dl K^{ab} \aeq \ep^a_{\ c}K^{cb}+\ep^{b}_{\ c}K^{ac}.
}
Here, $\ep^{ab}$ are infinitesimal parameters which are functions of the space-time points and satisfy $\ep^{ab}=-\ep^{ba}$. 
We put
\bea{
\eta^a =\ast \theta^a,\ \eta^{ab}=\ast(\theta^a \w \theta^b), \ \eta^{abc}=\ast(\theta^a \w \theta^b \w \theta^c), \ \eta^{abcd}=\ast(\theta^a \w \theta^b \w \theta^c \w \theta^d) .
}
In Appendix \ref{formula}, several identities about 
$\theta^b \w \eta_{a_1 \cdots a_r}$ $(r=1,2,3,4)$ , $\dl \eta_{a_1 \cdots a_r}(r=0,1,2,3)$ and $d\eta_{a_1 \cdots a_r}$ $(r=1,2,3)$  are listed.

The Lagrangian form of the gravity in the second order formalism is given by
\bea{
L(\theta,d\theta) = L_\G(\theta,d\theta)+L_\RM{mat}(\theta,d\theta, \psi, d\psi).
}
Here, $L_\G$ is the Lagrangian form for the pure gravity given by \cite{K, 2015}
\bea{
L_\G(\theta,d\theta) = \f{1}{2\ka}W \com W := \Om^{ab} \w \eta_{ab} -d(\om^{ab} \w \eta_{ab}) \la{def_N^pr},
}
and $L_\RM{mat}(\theta,d\theta, \psi, d\psi):=L_\RM{mat}(\theta,\om(\theta,d\theta), \psi, d\psi)$ is the Lagrangian form of \co{matters} $\psi$ which are scalar fields, Dirac fields and gauge fields.
Here, $\ka$ is the Einstein constant. 
Only the Dirac fields couple to $\om^{ab}$.

The local Noether current $N$ for the transformation \re{trans_gravity} is given by
\bea{
N \aeq \half \ep^{ab} N_{ab}+\half d\ep^{ab} \w F_{ab},
}
with
\bea{
N_{ab} \aeq N_{ab}^{(0)} +N_{ab}^{(1)}, \\
N_{ab}^{(0)} \aeq 2\f{\p L_\RM{mat}}{\p \om^{ab}} ,\la{N_ab^1}\\
N_{ab}^{(1)} \aeq \theta_{b}\w \pi_{a}-\theta_{a}\w \pi_{b}, \\
F_{ab} \aeq -\f{1}{\ka}\eta_{ab} .\la{F_ab}
}
The global Noether currents $N_{ab}^{(1)}$ satisfy
\bea{
\{N_{ab}^{(1)}, N_{cd}^{(1)} \}\aeq \g_{bc}N_{ad}^{(1)}-\g_{ac}N_{bd}^{(1)}+\g_{ad}N_{bc}^{(1)}-\g_{bd}N_{ac}^{(1)}. \la{G_LG}
}
We derive \re{F_ab} in Appendix \ref{A_derivation}.
We can confirm that $N$ is the generator of the local Lorentz transformation:
\bea{
\dl \theta^a \aeq \{\theta^a, N \} = \ep^a_{\ b}\theta^b ,\\
\dl \pi_a \aeq \{\pi_a, N \}= -\ep^{b}_{\ a}\pi_{b}-d\ep^{bc} \w \f{1}{2\ka} \eta_{abc}.
}

\section{Summary}
In the covariant canonical formalism (CCF), 
we showed that the Noether current is the generator of the infinitesimal local transformation of $p$-form dynamical fields $\psi^A \to \psi^A +\dl \psi^A$
if the transformation of the Lagrangian form is given by $\dl L=dl$ and $\dl \psi^A$ and $l$ depend on only $\psi^A$ and the parameters. 
As an instance, we studied the local gauge transformation for the gauge field and the local Lorentz transformation for the second order formalism of gravity. 
Our demonstrations are for non-constrained systems in the CCF, in contrast to previous studies \cite{2019, 2020} for constraint systems.

\appendix

\section{\rd{Derivation of \re{pb}}} \la{A_PB}

Using \re{PB}, the left-hand side of \re{pb} is given by
\bea{
\{G_r , G_s \} \aeq  \f{\p G_r}{\p \psi^A}\w \f{\p G_s}{\p \pi_A}-(-1)^{(d+p-1)d}\f{\p G_r}{\p \pi_A}\w \f{\p G_s}{\p \psi^A}
+\f{\p G_r}{\p A^t}\w \f{\p G_s}{\p \pi_t}-(-1)^d \f{\p G_r}{\p \pi_t}\w \f{\p G_s}{\p A^t}.
}
Using
\bea{
\f{\p N_r^{(0)}}{\p \psi^A} =(\bm{G}_r)^B_{\ A}\pi_B ,&\ \  \f{\p N_r^{(0)}}{\p \pi_A} = (-1)^{p(d-p-1)}(\bm{G}_r)^A_{\ B}\psi^B ,\\
\f{\p N_r^{(1)}}{\p A^t} = f^s_{\ rt}\pi_s ,&\ \ \f{\p N_r^{(1)}}{\p \pi_t} = (-1)^d f^t_{\ rs}A^s ,
}
we obtain
\bea{
\{N_r^{(0)} , N_s^{(0)} \} \aeq  (\bm{G}_r)^B_{\ A}(\bm{G}_s)^A_{\ C} (-1)^{p(d-p-1)} \pi_B \w \psi^C
- (\bm{G}_s)^B_{\ A} (\bm{G}_r)^A_{\ C}  \psi^C \w \pi_B \no\\
\aeq ([\bm{G}_r, \bm{G}_s])^A_{\ B} \psi^B \w \pi_A \no\\
\aeq f^t_{\ rs}N_t^{(0)}
}
and
\bea{
\{N_r^{(1)} , N_s^{(1)} \} \aeq f^t_{\ rs}N_t^{(1)}. \la{pb_N_1}
}
For \re{pb_N_1}, we used the Jacobi identity $f^p_{\ rs}f^q_{\ pt}+f^p_{\ st}f^q_{\ pr}+f^p_{\ tr}f^q_{\ ps}=0 $.
Because of $\{N_r^{(0)} , N_s^{(1)} \}=0=\{N_r^{(1)} , N_s^{(0)} \}$, we obtain \re{pb}.

\section{\rd{Covariant canonical formalism for gauge filed}} \la{YMU}

The Lagrangian form of the gauge field is given by
\bea{
L_1 = -\f{1}{2k} \mF^r \w \ast \mF_r .\la{L_gauge}
}
Here,  $k$ is a positive constant, $\mF^r := dA^r +\half f^r_{\ bc}A^b \w A^c$ is the curvature of the gauge field, 
and $\mF_r :=\ka_{rs}\mF^s$ where $\ka_{rs}:= -f^a_{\ rb}f^b_{\ sa}(=\ka_{sr})$ is the Killing form.
The conjugate form of $A^a$ is given by
\bea{
\pi_a=-\f{1}{k}\ast \mF_a.
}
Because $dA^a$ can be represented by the conjugate form $\pi_a$ as $dA^a = k \ast \pi^a-\f{1}{2} f^a_{\ bc}A^b \w A^c$, 
the gauge field is a non-constrained system in the CCF. 
Here, $\pi^a := (\ka^{-1})^{ab}\pi_b= -\f{1}{k}\ast \mF^a$. 
All components of $\pi_a$ are independent. 
The number of components of $\pi_r$ is not equal to the number of components of $A^r$; $d(d-1)/2 \ne d$. 
The Hamilton form is given by
\bea{
H\aeq -\f{1}{2} f^a_{\ bc} A^b \w A^c\w \pi_a+\f{k}{2} \pi_a \w \ast \pi^a -L_0(\psi^A,(D\psi)^A). 
}
The derivatives of $H$ are given by
\bea{
\f{\p H}{\p A^a} \aeq -f^c_{\ ab}  A^b \w \pi_c-J_a ,\la{p_H_1}\\ 
\f{\p H}{\p \pi_a} \aeq k \ast \pi^a-\f{1}{2} f^a_{\ bc}A^b \w A^c. 
}
Here, 
\bea{
J_r := \f{\p L_0(\psi^A,(D\psi)^A)}{\p A^r}= (\bm{G}_r)^A_{\ B}\psi^B \w \pi_A=N_r^{(0)}. \la{J=N}
}
The canonical equations $dA^a=\p H/\p \pi_a$ and $d\pi_a = \p H/\p A^a$ become 
\bea{
dA^a \aeq k \ast \pi^a-\f{1}{2} f^a_{\ bc}A^b \w A^c,\\
d\pi_a \aeq -f^c_{\ ab}  A^b \w \pi_c-J_a .
}
The former is equivalent to the definition of $\pi_a$.
The latter is  equivalent to the Yang-Mills-Utiyama equation.
The covariant canonical formalism does not need gauge fixing. 

\section{Formulas} \la{formula}

Several useful formulas are listed. 
For $\theta^b \w \eta_{a_1 \cdots a_r}(r=1,2,3,4)$, 
\bea{
\theta^b \w \eta_{a_1\cdots a_r} \aeq (-1)^{r-1} r \dl^b_{[a_1}\eta_{a_2 \cdots a_r]} \la{A0}
}
hold \cite{95}. $[$ $]$ denotes the antisymmetrization. 
Using this, we have
\bea{
\theta^a \w \theta^b \w \eta_{cd} \aeq (\dl^a_c\dl^b_d-\dl^a_d\dl^b_c)\eta. \la{A4} 
}
For $\dl \eta_{a_1 \cdots a_r}(r=0,1,2,3)$, 
\bea{
\dl \eta_{a_1\cdots a_r} \aeq \dl \theta^b \w \eta_{a_1\cdots a_r b}  \la{A01} 
}
hold.
For $d\eta_{a_1 \cdots a_r}(r=1,2,3)$,
\bea{
d \eta_{a_1\cdots a_r} \aeq (r+1)A^b_{\ [a_1}\w \eta_{ba_2 \cdots a_r]} \no\\
\aeq (r+1)\om^b_{\ [a_1}\w \eta_{ba_2 \cdots a_r]}+ \Theta^b \w \eta_{a_1\cdots a_r b} \la{A8} 
}
hold.

\section{Derivation of \re{F_ab}} \la{A_derivation}

We derive \re{F_ab}.
$W$ of \re{def_N^pr} can be rewritten as
\bea{
W \aeq W^\ast + \mK \com 
W^\ast := A^a_{\ c}\w A^{cb} \w \eta_{ba} ,
}
with $\mK := K^a_{\ c}\w K^{cb} \w \eta_{ab}$. 
Then, by introducing $L_\G^\ast := W^\ast/(2\ka)$, we have
\bea{
\dl L \aeq \dl L_\G^\ast \no\\
\aeq \dl \theta^a \w \f{\p L_\G^\ast}{\p \theta^a}+d(\dl \theta^a) \w \f{\p L_\G^\ast}{\p d\theta^a} \no\\
\aeq \ep^a_{\ b}\Big (\theta^b \w \f{\p L_\G^\ast}{\p \theta^a}+d\theta^b \w p_a\Big)+ d\ep^{ab} \w \theta_{[b} \w p_{a]}, \la{dl_L_grav}
}
where \cite{K, 2015}
\bea{
p_a := \f{\p L_\G^\ast}{\p d\theta^a} =\f{1}{2\ka}A^{bc} \w \eta_{abc}.
}
Because of the global  Lorentz invariance, the first term of \re{dl_L_grav} vanishes.
Using \re{A8}, we have
\bea{
\theta_{[b} \w p_{a]} \aeq \f{1}{\ka}A_{[b}^{\ c} \w \eta_{a]c}=\half dF_{ab}.
}
Thus, we have
\bea{
\dl L \aeq  \half d\ep^{ab} \w dF_{ab} =dl \com l := -\half d\ep^{ab} \w F_{ab} .
}
The above equation and \re{N_current} lead to \re{F_ab}.

\section{\rd{Covariant canonical formalism of gravity}} \la{CCF_gravity}

The conjugate form of $\theta^a$ is given by \cite{2015}
\bea{
\pi_a=\f{1}{2\ka}\om^{bc} \w \eta_{abc}. \la{def_pi} 
}
The Hamilton form is given by \cite{2015}
\bea{
H =H_\G -L_\RM{mat} \com H_\G := \f{1}{2\ka} \om^a_{\ c}\w \om^{cb} \w \eta_{ba}.
}
The canonical equations are given by 
\bea{
d\theta^a \aeq  \f{\p H_\G}{\p \pi_a}+\Theta^a , \la{H1a} \\
d\pi_a \aeq \f{\p H_\G}{\p \theta^a}-\f{\p L_\RM{mat}(\theta,\pi)}{\p \theta^a}. \la{C_eq_Pi}
}
Here, we used $\f{\p L_\RM{mat}}{\p \pi_a}=-\Theta^a$ \cite{2015}.
$\Theta^a$ is the torsion 2-form.
Using \re{A4}, $ H_\G$ can be rewritten as
\bea{
 H_\G =  \f{1}{2\ka} (\om_{abc}\om^{bca}+\om_a \om^a)\eta.
}
Here, we expanded $\om_{ab}$ as $\om_{ab}=\om_{abc}\theta^c$ and put $\om_a := \om^b_{\ ab}$.
We can represent $\om_{abc}$ by $\theta^a$ and $\pi_a$ as
\bea{
\om_{abc} \aeq \ka \Big[ v_{c,ab}+\f{1}{d-2}(\g_{ac} v_b - \g_{bc} v_a) \Big] ,\la{om_v}\\
v_{c,ab} \aeqd -\ast V_{c,ab} \com V_{c,ab}:= \pi_c\w \theta_a \w \theta_b
}
with $v_a := v^{b}_{\ ab}$.
For an arbitrary $d$-form $\xi$, 
\bea{
\dl v_{c,ab} \xi \aeq - \dl v_{c,ab} \xi \ast \eta =- \dl v_{c,ab}\eta \ast \xi = (- \dl[ v_{c,ab}\eta]+ v_{c,ab}\dl \eta )\ast \xi \no\\
\aeq (-\dl V_{c,ab} + v_{c,ab}\dl \eta )\ast \xi
}
holds \cite{K}. 
Then, we have $\dl v_{c,ab} \eta = \dl V_{c,ab} - v_{c,ab}\dl \eta$.
Using this, we have
\bea{
\f{\p H_\RM{G}}{\p \pi_c} \aeq -\om^c_{\ a} \w \theta^a , \la{p_H_G} \\
 \f{\p H_\G}{\p \theta^d} \aeq  \f{1}{2\ka}(\om^{c}_{\ d} \w \om^{ab}\w \eta_{abc}+ \om^a_{\ c}\w \om^{bc} \w \eta_{bad}). \la{Goal}
}
Substituting \re{p_H_G} into \re{H1a}, we have
\bea{
d\theta^a = -\om^a_{\ b}\w \theta^b + \Theta^a ,\la{G_CE1}
}
which is equivalent to the first structure equation.
We can show that \cite{2015}
\bea{
-\f{\p L_\RM{mat}(\theta,\pi)}{\p \theta^c} = -T_c - \f{1}{2\ka}\om^{ab} \w\Theta^d \w \eta_{abcd} . \la{p_L_mat_pi}
}
Here, $T_c :=\p  L_\RM{mat}(\theta, \om)/\p \theta^c $. 
Substituting  \re{Goal} and \re{p_L_mat_pi} into \re{C_eq_Pi}, we have
\bea{
d\pi_c =  \f{1}{2\ka}(\om^d_{\ b} \w \om^{ab}\w \eta_{adc}+ \om^d_{\ c} \w \om^{ab}\w \eta_{abd}
 -\om^{ab}\w \Theta^d \w \eta_{abcd})-T_c . \la{G_CE2}
}

\end{document}